# Near-Zero Index Photonic Crystals with Directive Bound States in the Continuum

**Larissa Vertchenko[1,*], Clayton DeVault[2], Radu Malureanu[1], Eric Mazur[2], and Andrei Lavrinenko[1]**

[1]Department of Photonics Engineering, Technical University of Denmark, Ørsteds Plads 345V, DK-2800 Kongens Lyngby, Denmark

[2]Department of Physics and School of Engineering and Applied Sciences, Harvard University, 9 Oxford Street, Cambridge, MA 02138, United States of America

e-mail:lariv@fotonik.dtu.dk*, cdevault@g.harvard.edu, rmal@dtu.dk, mazur@seas.harvard.edu, alav@fotonik.dtu.dk

## ABSTRACT

Near-zero-index platforms arise as a new opportunity for light manipulation with boosting of optical nonlinearities, transmission properties in waveguides and constant phase distribution. In addition, they represent a solution to impedance mismatch faced in photonic circuitry offering several applications in quantum photonics, communication and sensing. However, their realization is limited to availability of materials that could exhibit such low-index. For materials used in the visible and near-infrared wavelengths, the intrinsic losses annihilate most of near-zero index properties. The design of all-dielectric photonic crystals with specific electromagnetic modes overcame the issue of intrinsic losses while showing effective mode index near-zero. Nonetheless, these modes strongly radiate to the surrounding environment, greatly limiting the devices applications. Here, we explore a novel all-dielectric photonic crystal structure that is able to sustain effective near-zero-index modes coupled to directive bound-states in the continuum in order to decrease radiative losses, opening extraordinary opportunities for radiation manipulation in nanophotonic circuits. Moreover, its relatively simple design and phase stability facilitates integration and reproducibility with other photonic components.

## Introduction

The increasing demand of computational power, efficiency and portability has boosted research in photonics engineering, where the study of materials and new ways to transmit information has become essential[1]. In particular, we see the rise of near-zero index materials and their influence on quantum technologies, promising to enable light manipulation for complex computational tasks[2,3]. Photonic chips are the predominant platform to create quantum computers, as well as sensors and communication devices[4,5]. The three main elements of such systems are sources of light, such as quantum dots, channels responsible for manipulating and transmitting information, and detectors in a phase-stable platform[6]. Each of these components must be designed taking into consideration their material, integration on-chip and functionality. An outstanding challenge regarding photonic chips is the existence of high impedance mismatch between its components, limiting integration and resulting in high reflections and phase deterioration. Hybrid integration approaches incorporate different photonic fabrication technologies, which would not be achieved by a single fabrication process[1]. In this way, complex on-chip systems could be realized integrating photon sources, waveguides and detectors.

One way to overcome the impedance mismatch is by using materials with the refractive index close to zero, the so-called near-zero index materials (NZIM)[7]. Inside these materials light will experience exotic phenomena, such as enlargement of the wavelength[8], enhancement of nonlinearities[9,10], high electric fields[11] and constant phase distribution[12]. In particular, the constant phase related to a small wave vector leads to important applications in quantum photonics[13,14]. A large number of optical components integrated on chip can introduce multiple phase changes. When photons interact with such environment they acquire random phase distributions that will disrupt their signal, causing decoherence, hence losing its quantum properties[15]. Inside NZIM, radiation will be coherent due to the constant phase property, enabling us to perform operations in reasonable time before information is lost. Moreover, low-index platforms exhibit impressive nonlinear efficiencies that would be otherwise achieved only by extremely intense laser sources. Its constant phase property also relaxes the phase matching condition for several nonlinear effects, such as harmonic generation processes. As most materials exhibit low nonlinearities, NZIM becomes essential for many applications that rely on nonlinear optical phenomena, including data storage and quantum information[9]. Our research aims to provide a phase-stable component that could be integrated to hybrid photonic chips, with low coupling losses due to small impedance difference.

Conductors such as transition nitrides, ITO and other transparent conductive oxides have been studied for NZI properties. However, their intrinsic losses hinder their use for waveguiding applications, at the visible and near-infrared wavelengths[16–18]. The relative permittivity ($\varepsilon$) and permeability ($\mu$) dictate how the charges of a medium will align to external electromagnetic fields and are related to the refractive index by $n = \sqrt{\mu\varepsilon}$. Thus, in principle these materials would also behave as NZIM. However, the imaginary part of $\varepsilon$, which is associated with intrinsic losses, is overly high annihilating near-zero index effects in the system[19]. As a consequence, researchers are exploring ways to use dielectrics instead of conductors to achieve near-zero regime without intrinsic losses[20,21]. It is well known that bulk dielectrics have refractive index above one. Periodically arranged dielectric structures are able to collectively excite resonances that might exhibit an effective mode index less than one. The Γ point lies at the center of the primitive cell in the reciprocal space (Brillouin zone), having a zero wave vector. By employing dielectric photonic crystals designed to support specific resonances[22] having accidental degeneracy at Γ, it is possible to attain an effective near-zero index mode[23]. For instance, Huang *et. al.*[24] showed that silicon pillars, of radius *r*, arranged periodically with lattice constant $a = 2r$ would exhibit two linear bands and an additional flat band intersecting at a triply degenerate point. For the electric filed oriented parallel to the cylinders axis, these three electromagnetic modes would correspond to two dipolar and a monopolar resonance. Close to the so-called Dirac's triple point the photonic crystal would be able to sustain an effective near-zero index, where both permittivity and permeability would be effectively near zero. In this way the wave impedance, given by $Z = \sqrt{\mu/\varepsilon}$, would match the free space, allowing complete integration between sources, waveguides and detectors[25,26]. Nonetheless, the modes excited in these structures are localized above the light-line, meaning that they easily couple to the environment, resulting in radiative losses.

Aiming to diminish leakage of radiation to the surroundings, several research groups[27–29] theoretically proposed near-zero index photonic crystals that could exhibit special resonances named bound states in the continuum (BICs)[30,31]. Two classes of BICs may be explored in a system[32]. Symmetry-protected BICs are a result of mismatch between the symmetry of modes in free space and the modes inside the system, occurring only at the Γ point[32,33]. Whereas trapped BICs are resonances in the system that collectively result in destructive interference outside the structure, preventing leakage of radiation[34,35]. Due to their nonradiative character, the Q-factor of BICs becomes infinite, meaning that radiation is completely confined in the system. In most BIC demonstrations, the out-of-plane absence of radiation is analyzed, while in-plane propagation happens in all allowed directions of the photonic crystal lattice. Nonetheless, there is a class of BIC, named separable BIC, where the direction of radiation can be manipulated in-plane, resulting in guided resonances while maintaining the out-of-plane fields inhibited. Its first confirmation was recently reported in a 1D photonic crystal trench[33]. Such guided resonances lead to important applications, such as control of radiation in energy efficient photonic crystal lasers and photonic antennas. It is an enormous challenge to combine BICs with a NZIM, mainly due to difficulties in finding feasible shapes of the photonic crystals' unit cell. Furthermore, NZIM are highly dependent on polarization, thus complicating the overall process of discovering the right geometry. In photonic circuitry, in-plane polarization (TE) is the preferable to work with, making the system of pillars previously suggested unfeasible for such polarization. Therefore, we investigated photonic crystal membranes made of silicon that would behave as NZIM and are able to contain separable BICs, at the near infrared range.

Here we report the experimental characterization of a novel Si photonic crystal geometry composed of circular holes of different sizes. In order to reach the NZI regime, we explored a $C_2$ rotational symmetry by hybridizing a square lattice with a triangular lattice, as depicted in Fig. 1a. By adjusting the structural parameters of the system, such as radius and period, we were able to achieve conditions for a trapped BIC in the ΓX direction. Theoretical calculation showed that such structure would be able to reach Q-factors as high as $10^{18}$, which is normally found by concentrating light in optical cavities. The system was designed to have a BIC and Dirac degeneracy near the wavelength of $\lambda = 1550$ nm. The $C_2$ symmetry applied to the lattice in the $x$ and $y$ directions resulted in the first observation of a separable quasi-BIC in a 2D photonic crystal, where radiation was able to propagate only in one direction inside the membrane (ΓX). We have performed transmission measurements leading to a Q-factor of 2546. Such discrepancy to theory can be explained by the finite size of the sample, limited equipment resolution and natural fabrication imperfections, which naturally break the symmetric character of BICs, resulting in high, but finite Q-factors[36]. Nonetheless, our result is still high when compared to trivial photonic crystal membranes[37]. Our achievements pave the way to novel photonic crystal devices with high nonlinearities, light confinement and constant phase distribution, as well as better integration due to near-zero index properties regarding impedance mismatch. Moreover, by exploring separable BICs we manage to have control over the light's propagation direction in-plane, finding important applications for optical switches and photonic circuitry.

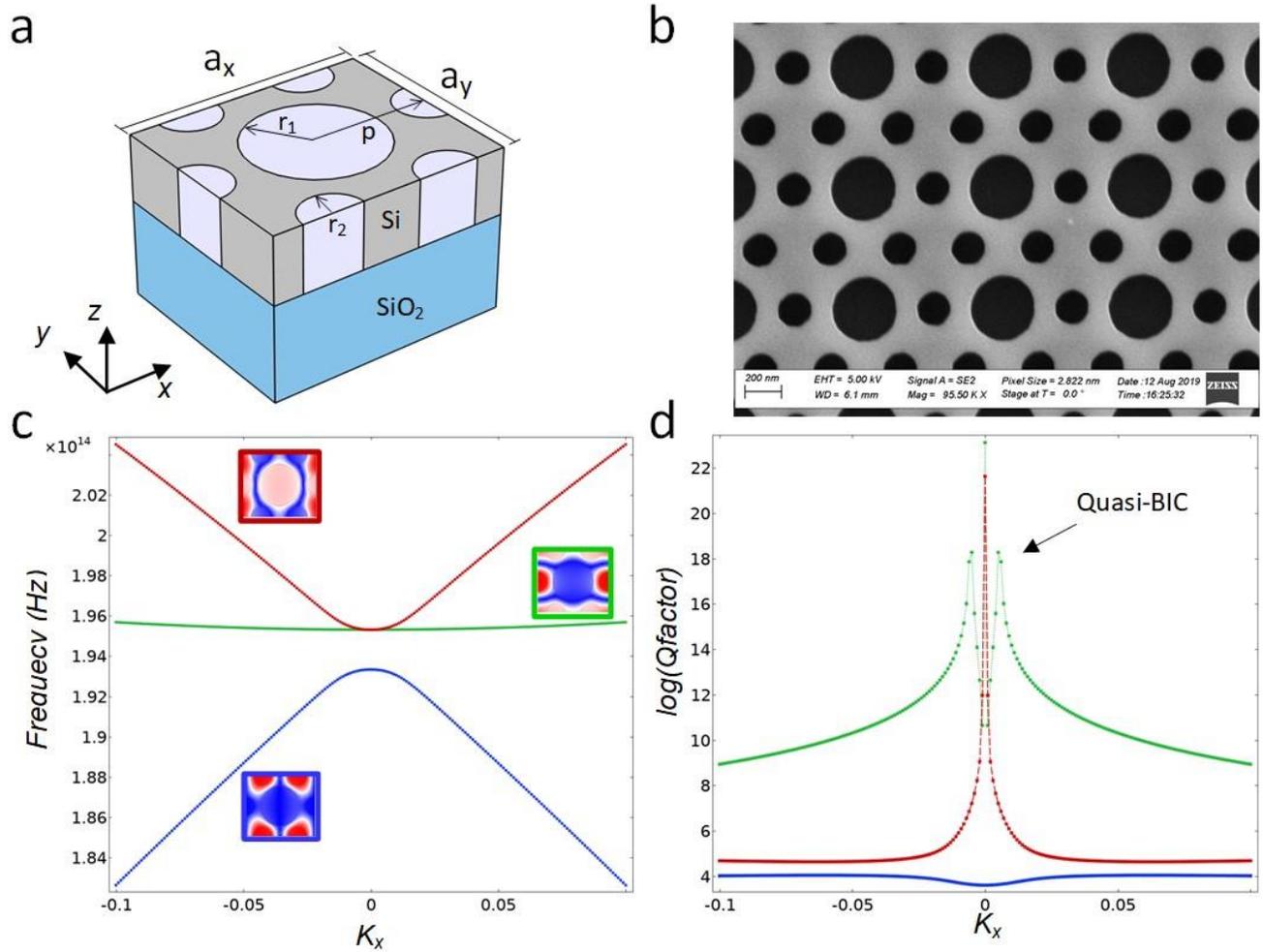

**Figure 1. Fabricated structure and theoretical results** a) Unit cell of the proposed design for a Si membrane on silica substrate, with air holes of radius 210 nm and 96 nm. b) SEM picture of a sample made of a 208 nm thick Si membrane on silica substrate. c) 2D band structure of the system in the ΓX where the color scheme indicates the electromagnetic modes in the membrane. d) Theoretical Q-factor for the proposed Si membrane with hybrid air holes calculated along $k_x$ through the equation $Q = \omega'/2\omega''$ .

## Results

### Material choice and fabrication approach

The fabricated structure consists of a 208 nm thick Si membrane with air holes of different radius, forming a flower-like shape pattern, as shown by the unit cell in Fig. 1a. An amorphous Si (a-Si) was deposited on fused silica glass followed by electron-beam lithography and resist development. The a-Si was etched using a Bosch process with the resist as a mask (see "Materials and methods"). The air holes were designed to have radius of 210 nm and 96 nm, as depicted in the scanning electron microscopy (SEM) image in Fig. 1b. The radial distance between the center of the biggest hole to the smaller ones is given by $p$ = 379 nm, which results in different lattice constants for the $x$ and $y$ directions ($a_x$ = 758 nm $a_y$ = 656 nm). Through numerical simulations using the commercially available software COMSOL[38], we were able to retrieve the band diagram as well as the electromagnetic modes, for TE polarization, of the designed structure, as shown in Fig. 1c. The Si was modelled with refractive index of 3.69, measured by ellipsometry for the wavelength of 1550 nm. A full 3D image of the band structure is presented in the Supplementary Information. From the dispersion diagram, we observed the same tendency for the combination of electromagnetic modes at the Dirac's triple point, where the color scheme indicates the different modes in the system. However, when the three modes intersect additional longitudinal waves are created due to radiation having electric field components aligned to the wave vector, disturbing the phase profile[39], which would be uniform in a homogeneous near-zero index material. In this case, we decided to isolate the lower frequency mode (quadrupole) that would be responsible for such fluctuations. At the wavelength of 1536 nm, two modes intersect representing what is called a Semi-Dirac point[40–42], which is created by the accidental degeneracy of two modes having a linear dispersion in one direction ($k_x$) and nonlinear relation in the perpendicular orientation ($k_y$). Close to $k = 0$, for the $k_x$ direction, the medium will have the effective permittivity and permeability simultaneously near-zero ($\varepsilon_{eff} \approx 0$ and $\mu_{eff} \approx 0$), while for $k_y$ it will behave as only an mu-near-zero (MNZ) material ($\varepsilon_{eff} \neq 0$ and $\mu_{eff} \approx 0$). These results give great possibilities to manipulate light propagation, such as optical switches, beam deflection and beam splitting[40]. So far, such system has not been experimentally reported in literature, being restricted to pillars working in the microwave range[43].

Due to the linear behavior near the Γ point in the $k_x$ direction, we were able to estimate the effective refractive index, $n_{eff}$ of the electromagnetic mode for $\lambda = 1536$ nm according to the relation $n_{eff} = ck/\omega$, achieving a value of $n_{eff} = 0.02$. We also demonstrate the existence of a trapped quasi-BIC by studying the tendency of the Q-factor in Fig. 1d, where it is possible to observe a local maximum displaced from Γ. By performing simulations of the photonic crystal with an eigenfrequency analysis[38], we may determine the corresponding eigenvalues $\Omega = \omega' + \omega''$ of the structure. Therefore, we calculated the Q-factor of the system following $Q = \frac{\omega'}{2\omega''}$, where its peak determines the existence of a quasi-BIC with extremely high Q-factors of order $10^{18}$. In principle, true BICs could be numerically achieved by tuning the holes dimensions with variation less than 1 nm.

### Experimental characterization

The optical characterization of the fabricated structure consisted of two parts, visualization of isofrequency contours at the Fourier plane and transmission measurements for Q-factor retrieval[33]. In both cases a tunable laser source (Santec TSL-550) with wavelength ranging from 1500 nm to 1630 nm was used, where a polarized laser beam is focused on the sample by a near-infrared 10X objective. A scheme of the setup is shown in Fig.2. In order to verify the wavelength of the near-zero index mode we analyzed the isofrequency contours in the reciprocal space, by using the resonance enhanced photon-scattering technique[44,45]. This approach relies on the fact that natural fabrication disorders scatter light not only at the resonance of interest, but also at different wave vectors close to it. All these resonances will radiate photons in the far field, which will then be projected into the Fourier plane by a 4f system, allowing us to reconstruct the isofrequency contour for various wave vectors. Since we need to interpret only the scattered signal that is directed towards the CCD camera, a second polarizer (P2) was used to block the reflected beam. In our setup, two 4f configurations (lenses L2-L5) were used to magnify the image onto the CCD. In particular lens L4 was used to switch from momentum space to real space, facilitating the alignment of the laser with the photonic crystal sample.

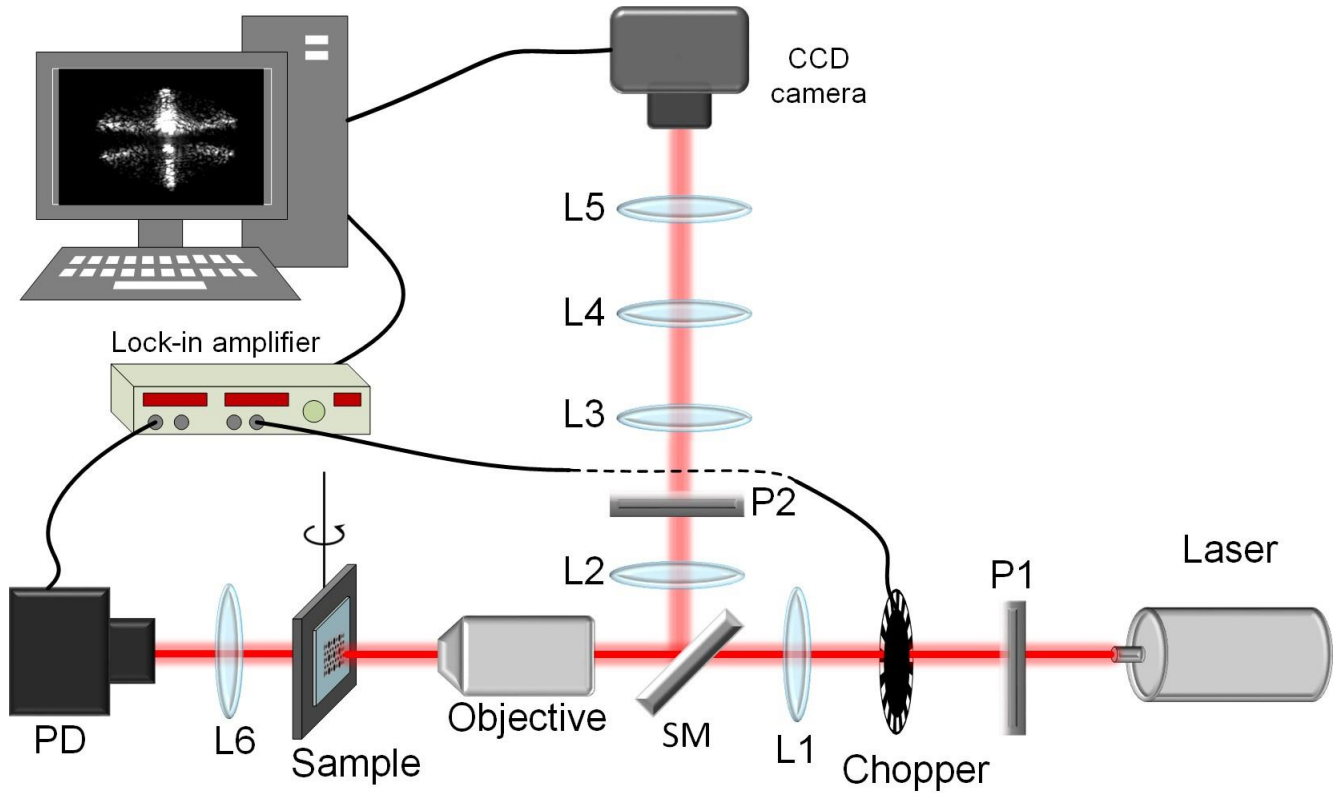

**Figure 2. Experimental setup** Light from a tunable laser source is focused on the sample by an NIR objective. L1 (focal length $f$ = +15 cm) is used to focus the beam onto the back focal plane of the objective. Lenses L2-L5 compose two 4f system to magnify and project the image into the Fourier plane. Polarizer P1 excites the right resonance in the sample, while P2 blocks the reflected signal so only scattered field reaches the CCD camera. For transmission measurements we use the phase sensitive detection, where a 1kHz chopper and the photodetector (PD) are connected to a lock in amplifier placed behind the sample. The full setup is integrated to a computer for data analysis.

By changing the laser's wavelength, we were able to visualize the isofrequency contour evolution and detect the wavelengths that corresponded to the degenerate points and quasi-BIC resonances. In Fig. 3a we make a comparison between the image captured by the CCD at the wavelength of the semi-Dirac point, 1536 nm, and theoretical simulations. We observed in the CCD picture three lines crossing, matching the theoretical prediction and proving the achievement of a NZIM. Apart from the semi-Dirac point wavelength, we were able to find two peculiar images of electromagnetic waves oriented along the $y$ direction at 1527.5 nm and 1551 nm, as shown in Fig. 3b. These field profiles are quasi-BICs, more precisely, they are called separable quasi-BICs[46], due to their unidirectional character. Since BICs are characterized by their high Q-factor, which is related to confinement, they cannot be inferred solely by looking at the band diagram. Thus, the transmission experiment becomes necessary in order to retrieve the Q-factor curve of the structure.

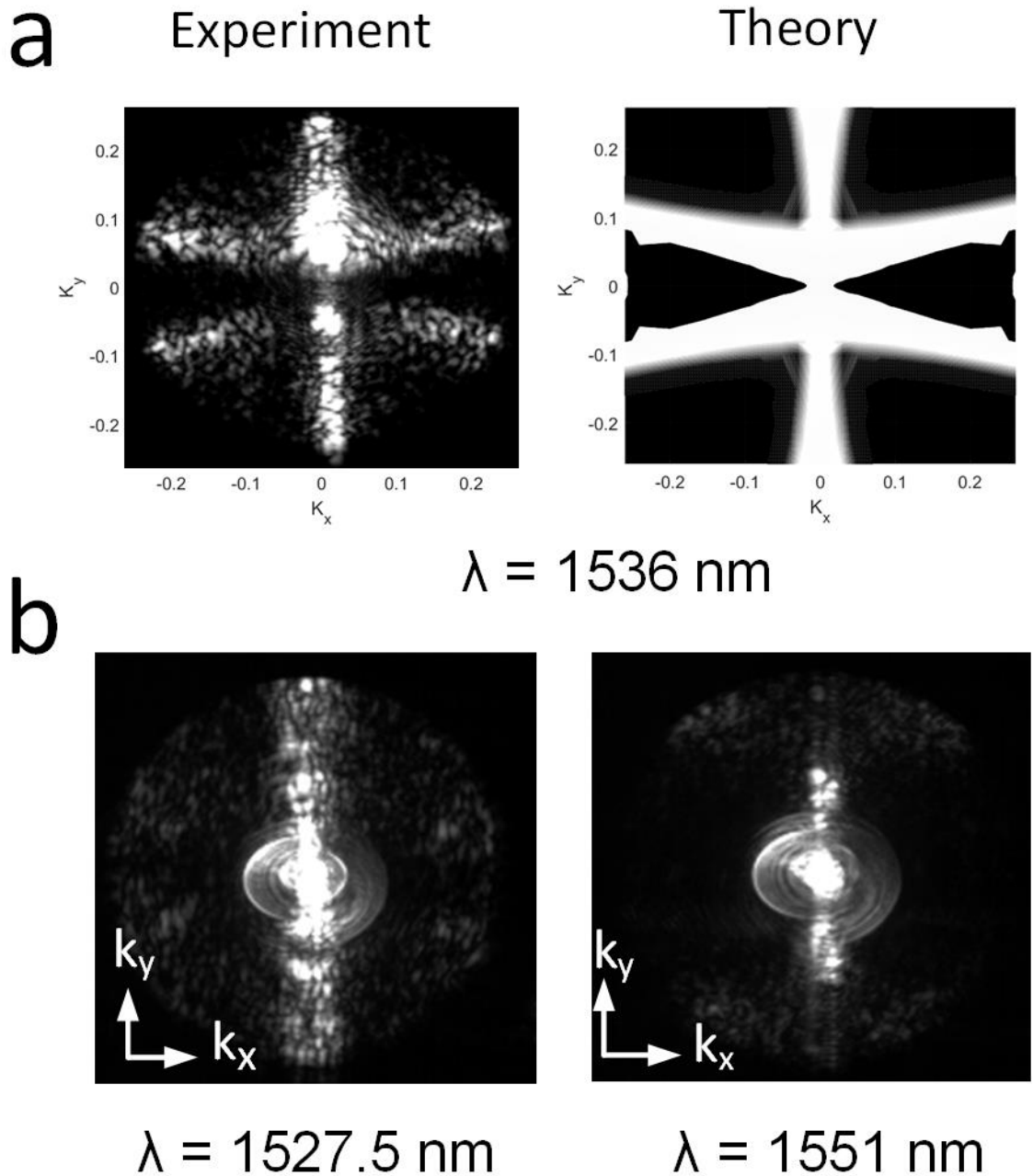

**Figure 3. Isofrequency images of a semi-Dirac point and separable quasi BICs** a) Isofrequency contour measured at the CCD camera (left) and obtained by numerical simulations (right), corresponding to the semi-Dirac point at the wavelength of 1536 nm. b) Isofrequency contour captured on the CCD camera for the separable quasi-BICs localized at the wavelengths 1527.5 nm 1551 nm.

When a system exhibits a BIC, the electromagnetic field is confined within at least one dimension. For instance, BICs in dielectric photonic crystal membranes will have vertical confinement. However, waves are still allowed to propagate in-plane, following the photonic crystal's band diagram. In our structure, the asymmetry applied in the photonic crystal lattice $(a_x \neq a_y)$ represents another constraint in the system where no BIC exists in the $k_y$ direction. In this case the separable BIC will propagate in a single direction inside the membrane, corresponding to the vertical straight lines we observed at the CCD in Fig. 3b, for two different wavelengths (1527.5 nm and 1551 nm). These scattered field patterns are evidence of the coupling between quasi-BICs and Bloch waves in the photonic crystal environment[47]. Their validation was done by performing transmission measurements at the second part of the experiment. In the setup, light transmitted through the sample reaches a photodector (Thorlabs DET10C2), which is connected to a lock-in amplifier synchronized with a 1-kHz chopper placed in the path of the incoming beam, for the phase-sensitive detection. The lock-in amplifier will acquire the signal from the chopper and use it as a reference to process the input signal from the sample, minimizing possible extra noise sources. The transmission spectrum was normalized to the one measured without the structure, as depicted in Fig. 4a, where it is possible to notice two thin dips highlighted in red. These dips represent the quasi-BICs around the semi-Dirac point wavelength, identified by the green line. Close to a quasi-BIC wavelength, narrow resonances (Fano resonances[48]) appear due to interference between the continuum resonant state and the discrete zero-index resonance. For each dip measured we have fitted a Fano resonance curve and retrieved its Q-factor, according to $Q = {}^{f}\!/_{\Delta f}$, where $f$ is the resonance frequency. The sample was attached to a rotation mount, where its angle was varied aiming to scan the q-factor for different wave vectors. The plots of the retrieved Q-factors as a function of the angle in respect to normal incidence are shown in Fig. 4b and Fig. 4c. The first quasi-BIC observed (Fig.4b) at 1527.5 nm is near the wavelength measured for the semi-Dirac point (1536 nm) and its Q-factor value peaks around 3 degrees, with value 2546. This means that the quasi-BIC will also behave as an NZI mode, at such wavelength. A second quasi-BIC was observed at 1551 nm (Fig.4c) with Q-factor peaking around 2500. It is important to emphasize that the measured Q-factors have lower values than the ones calculated from the simulations mainly due to fabrication imperfections and finite size of the sample. These values could be improved by making bigger samples or enhancing the system's symmetry by a full-suspended membrane, without any substrate.

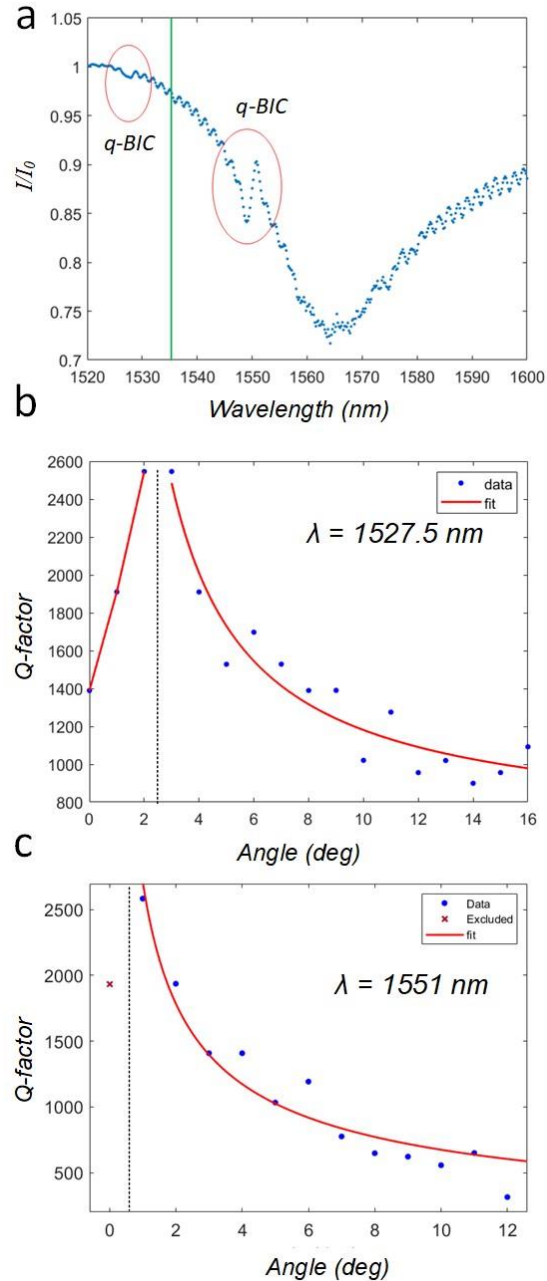

**Figure 4. Transmission and Q-factor measurements** a) Normalized transmission spectrum measured by the photodetector at normal incidence. The highlighted regions in red represent the first and second separable BICs, while the green line marks the Dirac's triple point wavelength (1536 nm).b) and c) are Q-factor plots as a function of the angles of the sample, where b) is the NZI quasi-BIC around 1527.5 nm and c) is another quasi-BIC related to dispersion. The red curves represent a rational fit, where the point for the lowest angle at c) was excluded due to lack of data for a proper fitting.

In order to interpret the existence of two quasi-BICs around the semi-Dirac point, we numerically analyzed the band diagram of the fabricated structure including the dispersion of Si, retrieved from ellipsometry measurements. The permittivity plots for the the Si slab may be found in the Supplementary Information. Although the modes excited in the system are still the same, the inclusion of dispersion has shifted the semi-Dirac point from $k_x = 0$ to $k_x = 0.004$, as shown in

Fig.5a. From this band structure, we have calculated the Q-factor, as depicted in Fig.5b. The results show that it is possible to observe two local maxima around the semi-Dirac point for the dipole mode, corresponding to the two quasi-BICs observed in the experiment. Deviations in wavelength from theoretical calculations are expected due to fabrication resolution and imperfections. For such reason, the semi-Dirac point, calculated at 1529 nm was measured at 1536 nm and the wavelength of the local maxima for the qBICs also varied.

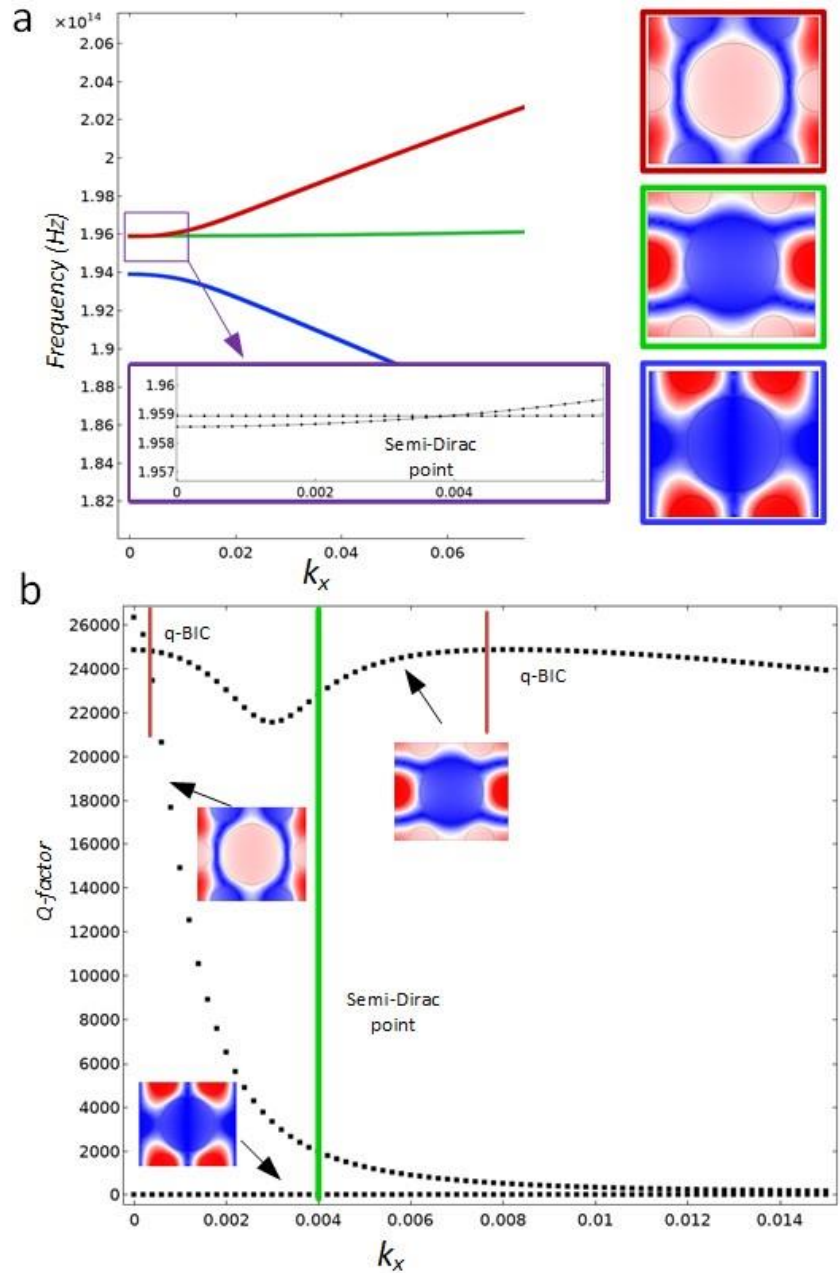

**Figure 5. Dispersion diagram and Q-factor of the fabricated structures** a) Band structure with the respective electromagnetic mode profiles, considering Si dispersion. The highlighted region shows the shift of the semi-Dirac point from the center of Brillouin zone. b) Q-factor retrieved from the band diagram, where local maxima (red lines) indicate quasi-BICs around the semi-Dirac point, located at the green line.

Since the degeneracy is no longer at the Γ point, the quadratic behavior of the dispersion becomes more pronounced imposing the question whether the effective parameters of the material remains near zero. In this sense we have validated the near-zero index behavior of the photonic crystal membrane by calculating its effective parameters using the boundary effective medium[49] approach and comparing our structures to a homogeneous medium. This method relies on evaluating the average eigenstate fields at the boundaries of the unit cell and its mathematical principle is explained in the Supplementary Information. The $C_2$ symmetry of the unit cell results in an anisotropic medium, observed by the difference in the field profile when the membrane is illuminated by a Gaussian beam with orthogonal polarization, as shown in Fig. 6a. Such system may be modelled as a homogeneous medium with effective parameters $\mu_{eff} = 0.0015$, $\varepsilon_{xeff} = 1.81$ and $\varepsilon_{yeff} = 0.16$, as depicted in Fig. 6b. The simulations were performed by considering dipole sources localized in each unit cell of the structure, where we have calculated the magnetic field profile ($H_z$). As a consequence of the finite size of the sample, edge effects will be more pronounced for the photonic crystal. In Fig. 6c we plot the transverse profile of $H_z$ in the XZ plane for a mode close the semi-Dirac point, at 1527 nm, where we notice a high confinement of the radiation, equivalent to a quasi-BIC regime. As expected from a near-zero index material, the phase of all elementary cells is constant leading to the wavefront shaping property, where an incoming electromagnetic wave is able to excite plane waves at the outgoing interfaces. Therefore, we were able to confirm the low-index property of our structure with effective refractive indexes equivalent to $n_{xeff} = 0.05$ and $n_{yeff} = 0.02$.

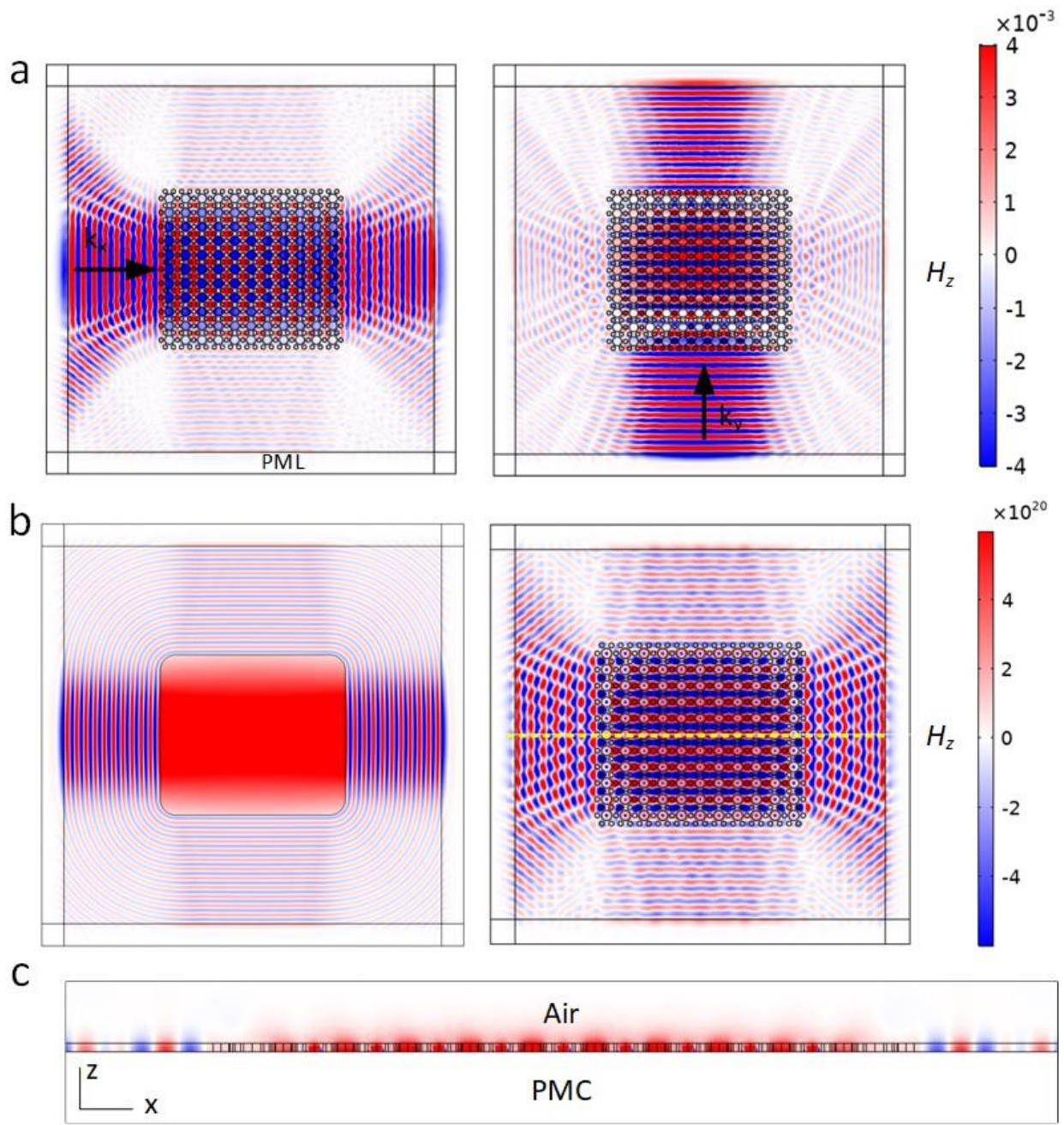

**Figure 6. Effective refractive index and field profiles** a) Magnetic field profile of the Si membrane illuminated by a Gaussian beam with orthogonal orientation, with wave vector indicated by the arrows. b) Comparison between a homogeneous structure (left) with effective parameters parameters $\mu_{eff} = 0.01$, $\varepsilon_{xeff} = 1.81$ and $\varepsilon_{yeff} = 0.16$ and the fabricated structure (right), where the electromagnetic modes were excited by dipole sources. c) Quasi-BIC at the wavelength of 1527 nm, where the transverse field profile was taken at the yellow dashed line. By using symmetry considerations for TE modes the perfect magnetic conductor was applied to decrease computational demand.

**Discussion**

We have experimentally verified the possibility to create a dielectric near-zero index material from a novel photonic crystal design of hybrid air holes in a Si membrane. The effective refractive index calculated for such structure was 0.02, which is unachieved by other platforms such as multilayers and homogeneous conductors, due to their intrinsic losses. By retrieving the isofrequency contour images of the structures, we were able to identify their effective NZI wavelength, corresponding to 1536 nm. Furthermore, we show that our structures are able to sustain a separable quasi-BIC helping to prevent radiative losses from the system while controlling the in-plane propagation direction of radiation. We have measured Q-factors as high as 2546 for the fabricated device. Although higher Q-factor values have been reported in literature for symmetric systems, we emphasize these results were obtained with glass as a substrate, which already deteriorates confinement. Better values could be reached by using self-standing membranes and a bigger area of the photonic crystal.

From the experimental results, we analyzed the effective constitutive parameters of the photonic crystal, validating its near-zero index property. Furthermore, the anisotropy of the quasi-BIC imposed by semi-Dirac cones due to $C_2$ rotation symmetry opens new possibilities for wave manipulation. Our hybrid structure of different circular holes greatly simplifies fabrication procedures. Such materials could pave the way for new structures in photonic crystal platforms, offering high electric field confinement, enhancement of nonlinearities and phase-stable environments.

**Materials and Methods**

*Fabrication process*

In order to obtain the structures with the desired characteristics, we have fabricated the samples in a class 100 cleanroom facility. In a first instance, amorphous Si (a-Si) was deposited on fused silica glass using LPCVD furnaces at 560degrees centigrade to ensure an amorphous and not a poly-crystalline structure of the deposited Si. The first deposition was used to determine the deposition rate and the refractive index using ellipsometry techniques. Once the deposition rate and the refractive index was known, the data was used for a quick numerical simulation to determine the thickness of the Si layer. We then performed a second deposition to obtain the desired thickness. After the second deposition, we measured the obtained thickness and refractive index. These final data were then used for a full simulation and readjustment of the structure's dimensions. We spun 150nm e-beam resist on the sample and then thermally deposited 20nm Al. This was done in order to avoid charging effects during electron-beam lithography. After the exposure, the Al layer was removed using standard wet-etching in $H_3PO_4 : H_2O$ solution. We have then proceeded with developing the resist. The a-Si was etched using a Bosch process with the resist as mask. The selectivity between the resist and the a-Si is 1:7.The remaining resist was then removed using an $O_2$ plasma. Since the LPCVD furnace deposits on both sides of the wafer, the last fabrication step was a dry isotropic etch of the bottom Si using $SF_6$ plasma.

*Numerical modelling*

We performed FEM simulations using the commercial software COMSOL Multiphysics. The structure was modelled using the unit cell of Fig. 1a) with Floquet periodic boundary conditions and perfectly matched layers at the top and bottom. The refractive index of Si used in the simulations was retrieved by ellipsometry measurements. Calculations of the band diagram and q-factors were performed by an eigenfrequency study.

**Acknowledgements**

This work was supported by the Danish National Research Foundation through NanoPhoton - Center for Nanophotonics, grant number DNRF147. R. M. acknowledges the support of DTU Nanolab for the fabrication of the structures. A. V. acknowledges the support from the Independent Research Fund Denmark, DFF Research Project 2 "PhotoHub" (8022-00387B), Villum Fonden. The Harvard University team acknowledges support from DARPA under contract URFAO: GR510802.

**Author contributions**

L.V. carried out the simulations and designed the samples. R. M. fabricated the samples. L.V, C.D and R.M. carried out the measurements. A. L. and E.M. supervised the research. All authors participated in the revision process of the manuscript.

**Competing interests**

The authors declare that they have no conflict of interest